\documentclass[aps,prl,groupedaddress,fleqn]{revtex4}
\pdfoutput=1
\usepackage{graphicx}
\usepackage{amsmath}
\begin{document}

\title{Varying vacuum energy of a self-interacting scalar field}
\author{K. Trachenko$^{1}$}
\address{$^1$ School of Physics and Astronomy, Queen Mary University of London, Mile End Road, London, E1 4NS, UK, k.trachenko@qmul.ac.uk}

\begin{abstract}
Understanding mechanisms capable of altering the vacuum energy is currently of interest in field theories and cosmology. We consider an interacting scalar field and show that the vacuum energy naturally takes any value between its maximum and zero because interaction affects the number of operating field modes, the assertion that involves no assumptions or postulates. The mechanism is similar to the recently discussed temperature evolution of collective modes in liquids. The cosmological implication concerns the evolution of scalar field $\phi$ during the inflation of the Universe. $\phi$ starts with all field modes operating and maximal vacuum energy in the early inflation-dominated epoch. As a result of inflation, $\phi$ undergoes a dynamical crossover and arrives in the state with one long-wavelength longitudinal mode and small positive vacuum energy predicted to be asymptotically decreasing to zero in the late epoch. Accordingly, we predict that the currently observed cosmological constant will decrease in the future, and comment on the possibility of a cyclic Universe.
\end{abstract}

\maketitle

\section{Introduction}

The vacuum energy of the quantized field, $E_0$, is the sum of ground-state energies of harmonic oscillators calculated to be $E_0\propto\hbar k_{\rm max}^4$, where $k_{\rm max}$ corresponds to the maximal energy where the field theory applies, tentatively at the Planck scale \cite{carroll}. In this consideration, $E_0$ is a static quantity governed by $k_{\rm max}$ only. Here, we point out that $E_0$ of the field interacting in the form of a double-well potential is a dynamical quantity which can take any value between its maximum and zero depending on the state of the system. This takes place because interaction affects the number of field modes at operation.

In a simplified case where the cosmological constant $\Lambda$ is related to the vacuum energy of the scalar field \cite{carroll,linde,liddle}, our result has implications for the cosmological constant problem \cite{carroll,adams,linde,liddle,bass,silve,sahni,weinberg,martin,sola-review}. Here, a major challenge is seen in uncovering an underlying physical structure that governs the currently observed small cosmological constant and its evolution in different epochs \cite{carroll}. Quantitatively, calculating the vacuum energy as $E_0\propto\hbar k_{\rm max}^4$ returns the value vastly larger than the observed cosmological constant. The important question is what physical mechanism can alter the vacuum energy of the field, with the scalar field being a commonly considered case \cite{carroll,adams,linde,liddle}?

This question is the focus of our paper where we propose a model predicting small positive vacuum energy and its future evolution. The cosmological scalar field starts with all field modes operating and maximal vacuum energy in the early inflation-dominated epoch and, as a result of inflation, undergoes a dynamical crossover and arrives in the state with one long-wavelength longitudinal mode and small positive vacuum energy predicted to be asymptotically decreasing to zero in the late epoch. We predict that the currently observed cosmological constant will decrease in the future, and comment on the possibility of a cyclic Universe.

Dynamically evolving energy components of the Universe were discussed before (see, for example, Refs. \cite{sahni,stein,carvalo,basilakos,sola1,sola2,sola3} and references therein). In these discussions, the dynamical component is postulated in order to account for the experimental data, or assumptions are made about the component's dynamics. Here, we find that varying vacuum energy and its possible smallness are natural consequences of evolving field modes as a result of field self-interaction, as is the case for liquid modes. This assertion follows once a double-well form of the interaction potential is considered and involves no further assumptions. This is a new result that has not been hitherto anticipated or discussed.

In our analysis, we discuss the direct analogy between liquids and interacting fields. The analogy is interesting to follow though it is not required to derive the main result of how the vacuum energy becomes small. In the next chapter, we summarize our recent theory of liquids based on collective modes. The summary presents a physical mechanism of how collective modes evolve in a strongly-interacting system. This is followed by the discussion of the mechanism of varying the vacuum energy of the field and cosmological implications.

\section{Evolution of collective modes in liquids}

We consider the double-well potential in Figure 1 that endows atoms in liquids with the ability to jump between neighbouring quasi-equilibrium positions. First-principles treatment of this process is not tractable due to strong interactions and their non-linearity \cite{annals}, however the problem was solved by J Frenkel \cite{frenkel} using what we call the ``Frenkel reduction''. Frenkel introduced $\tau$ as the average time between two consecutive atomic jumps in the liquid, or two adjacent minima of the potential in Figure 1. He observed that at short time $t<\tau$, the system is a solid and therefore supports all three collective modes: one longitudinal and two transverse modes. At long time $t>\tau$, the system is a liquid where no transverse modes exist because the equations of motion cease to be solid-like. This means that the liquid supports transverse modes with frequency above $\omega_{\rm F}=\frac{1}{\tau}$ only:

\begin{equation}
\omega>\omega_{\rm F}=\frac{1}{\tau}
\label{tau}
\end{equation}
\noindent but not below. Here, $\omega_{\rm F}$ is the Frenkel hopping frequency.

The longitudinal mode is unaffected in this regime (apart from different dissipation laws) because density fluctuations are present in any elastic medium, and exist up to the shortest wavelength if the system is dense enough \cite{prb1,col-review}.

\begin{figure}
\begin{center}
{\scalebox{0.35}{\includegraphics{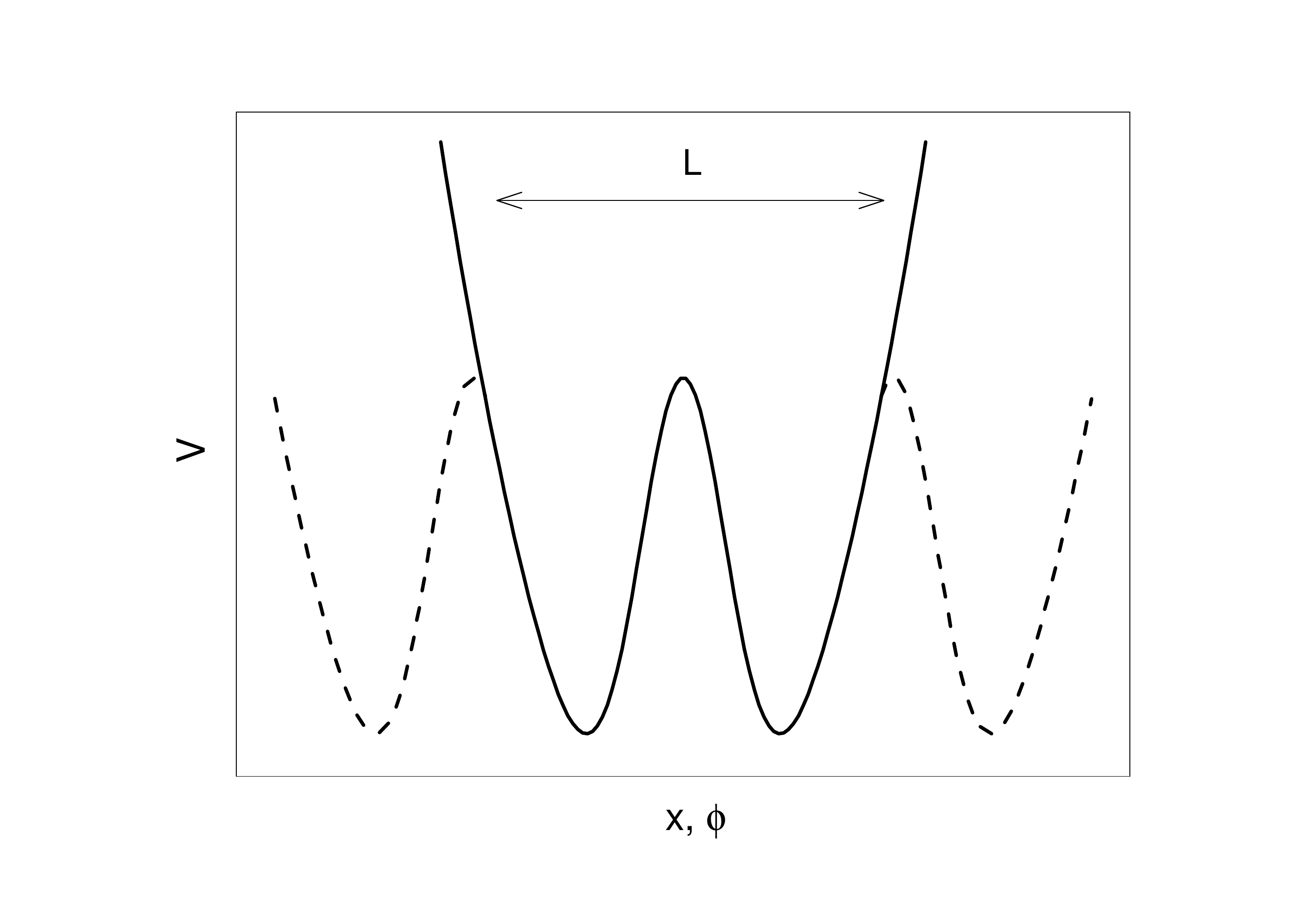}}}
\end{center}
\caption{Double-well (or multi-well) potential. The horizontal axis represents atomic coordinate $x$ in the liquid. The horizontal axis also represents the normal mode $\phi$ in the field Hamiltonian (\ref{field}), because the presence of the double-well interaction in the field Hamiltonian gives the energy spectrum of the liquid as discussed between Eq. (\ref{field}) and Eq. (\ref{phi1}) in the text.}
\label{wells}
\end{figure}

The ability of liquids to support high-frequency transverse modes has been verified experimentally fairly recently, and with a significant time lag after Frenkel's prediction \cite{col-review}. We have recently used this ability to construct the phonon theory of liquid thermodynamics consistent with the experimental data on liquid heat capacity \cite{prb1,col-review,scirep1}. In this theory, liquid energy includes the contribution of one longitudinal mode and two transverse modes with spectrum given in Eq. (1). Generally, $\tau$ decreases with temperature (and increases with pressure). Then, the number of transverse modes in the liquid becomes progressively smaller with temperature, according to Eq. (1). This explains the experimental decrease of liquid specific heat from $3k_{\rm B}$ to $2k_{\rm B}$ at which point $\omega_{\rm F}$ reaches its maximal Debye frequency, and the system loses all of its transverse modes \cite{prb1,col-review,scirep1}.

Progressive reduction and eventual disappearance of transverse waves in the liquid according to Eq. (1) is the {\it first} mechanism of how collective modes evolve in the system as a result of the double-well interaction in Figure 1. The {\it second} mechanism is the reduction and disappearance of the remaining longitudinal mode. This takes place above the Frenkel line on the phase diagram \cite{phystoday,prl}. The line separates the combined oscillatory-diffusive motion in the liquid-like regime at low temperature and purely diffusive motion in the gas-like regime at high, the change of dynamics particularly apparent in the supercritical state where no phase transition intervenes. Exactly at the Frenkel line, the system has one longitudinal mode with wavelengths spanning from the shortest interatomic separation up to the system size. Further temperature increase in the diffusive regime above the line increases particle mean free path $L$ (see Figure 1), the average distance between consecutive collisions. This removes longitudinal modes with wavelengths smaller than $L$ because the motion is diffusive and not oscillatory at those lengths. Hence, the remaining longitudinal modes start with wavelengths larger than $L$:

\begin{equation}
\lambda_l>L
\label{length}
\end{equation}

Temperature increases $L$ by elevating the system further above the activation barrier. Pressure decreases $L$ by effectively increasing the activation barrier and hence lowering the system closer to the liquid-like hopping regime.

Recently, we have shown that the energy of these modes is consistent with experimental specific heats of supercritical fluids. In the gas-like regime of particle dynamics, the specific heat decreases from $2k_{\rm B}$ to $\frac{3}{2}k_{\rm B}$, the ideal-gas value corresponding to the absence of collective modes \cite{col-review,natcom}.

We note that the collective modes under discussion include high-frequency modes that make the largest contribution to the liquid energy and whose evolution governs liquid thermodynamics \cite{col-review}. These modes propagate in the solid-like elastic regime $\omega\tau>1$ (see (\ref{tau}) and Refs. \cite{prb1,col-review,scirep1}). The solid-like regime contrasts the commonly considered hydrodynamic regime of liquids where $\omega\tau<1$. The same applies to the field theory discussed below: we will consider high-energy field modes not describable by the hydrodynamic approach.

\section{Vacuum energy of a self-interacting field}

We now discuss the evolution of the vacuum energy of the self-interacting scalar field $\phi$ and consider a Hamiltonian

\begin{equation}
H=\frac{1}{2}\sum_i\left(\dot{\phi_i}^2+\omega_i^2\phi_i^2+V(\phi_i)\right)
\label{field}
\end{equation}

In (\ref{field}), the first two terms represent independent harmonic modes with frequencies $\omega_i$ as is the case for, for example, Klein-Gordon field. Our harmonic (non-interacting) Hamiltonian has physically sensible positive $\omega_i^2$, in contrast to the common consideration starting with the imaginary masses (frequences) and subsequent adding the $\phi^4$ term in order to stabilize the field and for the initially massless mode to acquire mass via the Higgs effect \cite{bogol}.

For simplicity and definitiveness but without the loss of the generality, we consider a real scalar field $\phi$ with three components $\phi^1,\phi^2,\phi^3$, similar to normal modes of the phonon field, although the number of components can be larger. The model has a rotational symmetry. The interaction term $V(\phi)$ has the double-well form shown in Figure \ref{wells}. A complex scalar field of Higgs type can also be discussed as in our previous consideration where $V(\phi)$ depends on even powers of $\lvert\phi\rvert^2$ and has the ``Mexican hat'' form \cite{scirep3} with a continuous family of global minima around a potential barrier. In this case, the double-well potential is the two-dimensional cut of the hat by the vertical plane. As far as the evolution of field modes and the mechanism of smallness of the vacuum energy are concerned, considering real or complex fields gives the same result: small vacuum energy is related to the regime of gas-like dynamics above the potential barrier.

It is interesting to ask about the origin of the double-well (or multi-well) potential in Figure \ref{wells} and how generic it is. In condensed matter, the double-well form originates from a finite energy of particle interaction. This energy sets the characteristic height of the activation energy barrier separating two minima of the potential. Particle motion in the resulting double-well (or multi-well) potential gives three different regimes of particle dynamics and three states of matter: solids, liquids and gases. Particles oscillate in one single minimum in solids, oscillate and diffusively move between different minima in liquids and diffusively move above the potential barrier in gases. If $K$ and $P$ are kinetic and potential energy of the system, $K\ll P$, $K\approx P$ and $K\gg P$ give solid, liquid and gas state, respectively. One can assume that the same applies to interaction between field modes, resulting in the double-well interaction potential and different regimes of field dynamics.

\subsection{Large vacuum energy in the liquid-like regime}

We start with the {\it first}, liquid-like, regime of field dynamics. Recently, we have shown that Hamiltonian (\ref{field}) is identical to the Hamiltonian describing the liquid if $\phi$ is the liquid normal mode in the sense that both Hamiltonians yield the same energy spectra \cite{scirep3}. Indeed, if, for example, the interaction is taken as $V(\phi_i)=-\frac{g}{2}\phi_i^4+\frac{\lambda}{6}\phi_i^6$, the total potential, $\frac{1}{2}\omega_i^2\phi_i^2-\frac{g}{2}\phi_i^4+\frac{\lambda}{6}\phi_i^6$, has one minimum if $\omega_i>\omega_{\rm F}=\sqrt{\frac{g^2}{\lambda}}$, and two minima on the ``Mexican hat'' if $\omega_i<\omega_{\rm F}=\sqrt{\frac{g^2}{\lambda}}$. In the latter case, symmetry breaking due to hopping of $\phi$ (via tunneling or temperature activation) results, according to the Goldstone theorem, in two modes removed and one mode remaining. Hence, the presence of double-well interaction in mode Hamiltonian (\ref{field}) results in one mode unmodified (``longitudinal'' mode) and two modes propagating only above a certain frequency $\omega_{\rm F}$ (``transverse'' modes), i.e. exactly the energy spectrum of liquids discussed by Frenkel. Below we continue to refer to these field modes as longitudinal and transverse using the analogy with liquids, although the terms do not carry the same meaning as in condensed matter.

Therefore, the Hamiltonian (\ref{field}) of the field dynamics in the low-temperature ``liquid-like'' regime reads:

\begin{equation}
\begin{aligned}
H=\frac{1}{2}\left(\sum_i\left(\dot{\phi_i^1}\right)^2+\sum_i\omega_i^2\left(\phi_i^1\right)^2\right)+\\
\frac{1}{2}\left(\sum_i\left(\dot{\phi_i}^{2,3}\right)^2+\sum_{\omega_i>\omega_{\rm F}}\omega_i^2\left(\phi_i^{2,3}\right)^2\right)
\label{phi1}
\end{aligned}
\end{equation}
\noindent where subscript 1 corresponds to the longitudinal mode and subscripts 2 and 3 to two transverse modes.

According to Eq. (\ref{phi1}), the modes at operation include one longitudinal mode and two transverse modes above frequency $\omega_{\rm F}$, i.e. exactly the modes operative in the liquid below the Frenkel line discussed above. The identity of Hamiltonians of liquids and fields interacting with potential in Figure 1 implies that their energy spectra and other properties are identical \cite{annals}.

We note in passing that Eq. (\ref{phi1}) implies the energy gap because the modes are summed not from zero as in the usual scheme of field quantization, but from the finite energy $\hbar\omega_{\rm F}$. This gives the mass gap and particles with mass $m_{\rm F}$: $m_{\rm F}=\frac{\hbar\omega_{\rm F}}{c^2}$ \cite{annals}.

Eq. (\ref{phi1}) applies to both classical and quantum fields. We now apply Eq. (\ref{phi1}) to the quantized field to calculate the zero-point energy of two transverse modes in (\ref{phi1}), $E_0^t$. Without the loss of generality of main results, $E_0^t$ can be evaluated in the simplified Debye-like model with quadratic density of states. In the liquid-like regime of field dynamics, $E_0^t$ can be evaluated as $\int\limits_{\omega_{\rm F}}^{\omega_{\rm max}}\frac{1}{2}\hbar\omega g_t(\omega)d\omega$, respectively, where $g_t=\frac{6\omega^2}{\omega_{\rm max}^3}$ is the Debye-like density of transverse states per mode and $\omega_{\rm max}$ is the maximal frequency, tentatively at the Planck scale \cite{prb1}. Similarly to liquids \cite{prb1,scirep1,natcom}, we assume that the dynamical modification of the energy spectrum in the above process (removing low-frequency transverse modes in the first dynamical mechanism and high-frequency longitudinal modes in the second mechanism) does not alter the remaining energy levels specified by Eq. (\ref{tau}) or Eq. (\ref{length}) and energy spacing. Hence, the density of states and energy spacing are normalized by $\omega_{\rm max}$ as in the non-interacting harmonic system.

Evaluating the integral gives $E_0^t$ per mode as

\begin{equation}
\begin{aligned}
&E_0^t=\frac{3}{4}\hbar\omega_{\rm max}\left(1-\left(\frac{\omega_{\rm F}}{\omega_{\rm max}}\right)^4\right), \omega_{\rm F}<\omega_{\rm max}\\
&E_0^t=\frac{3}{4}\hbar\omega_{\rm max}, \omega_{\rm F}=0
\label{zero1}
\end{aligned}
\end{equation}

In (\ref{zero1}), $E_0^t=\frac{3}{4}\hbar\omega_{\rm max}$ corresponds to the non-interacting field. In this case, $\omega_{\rm F}=0$, corresponding to field oscillations in a single harmonic well in Fig. (\ref{wells}).

In the liquid-like regime of field dynamics, the vacuum energy is maximal. For large $\tau$ (small $\omega_{\rm F}$), $E_0^t$ reaches its maximum: $E_0^t=\frac{3}{4}\hbar\omega_{\rm max}$ in Eq. (\ref{zero1}). This corresponds to many oscillations of the field $\phi$ in each well in Figure 1 and rare transitions between different wells. Adding the vacuum energy of the longitudinal mode $E_0^l=\frac{3}{8}\hbar\omega_{\rm max}$ (see Eq. \ref{zero2} below), the maximal vacuum energy is $E_0=E_0^t+E_0^l=\frac{9}{8}\hbar\omega_{\rm max}$.

\subsection{Small vacuum energy in the gas-like regime}

The {\it second} regime of field dynamics corresponds to the gas-like dynamics above the Frenkel line where two transverse modes disappear as discussed above, and the remaining excitation in the system is the longitudinal mode with the wavelength larger than $L$. Indeed, if $\phi$ in Hamiltonian (\ref{field}) is particle's coordinate, the collective excitations of this Hamiltonian in the gas-like regime are given by the longitudinal mode with the wavelength larger than $L$ as discussed above. Therefore, the Hamiltonian describing field dynamics in the gas-like regime is

\begin{equation}
H=\frac{1}{2}\left(\sum_i\left(\dot{\phi_i^1}\right)^2+\sum_{\lambda_i>L}\omega_i^2\left(\phi_i^1\right)^2\right)
\label{phi2}
\end{equation}

In (\ref{phi2}), $L$ is the mean free path of the field, and is equal to the longitudinal wavelength above which the dynamics of $\phi$ is oscillatory. Below $L$, the dynamics of the field is not oscillatory but is gas-like, corresponding to the Hamiltonian (\ref{phi2}) with zero second potential term and no modes. 
In the gas-like regime of field dynamics, the motion of the field is not bound to either of the wells in Figure 1, but becomes delocalized above the activation barrier where the restoring forces now operate at low frequencies (large wavelengths) determined by the outer walls of the potential, but not by the curvature of individual potential wells. %This regime is equivalent to the gas-like regime in liquids above the Frenkel line.

Eq. (\ref{phi2}) applies to both classical and quantum fields, similarly to Eq. (\ref{phi1}). We now apply Eq. (\ref{phi2}) to calculate the zero-point energy of field modes with the wavelengths larger than $L$.

In the gas-like regime of field dynamics, the vacuum energy $E_0^l$ can be evaluated as

\begin{equation}
E_0^l=\int\limits_0^{\omega_L}\frac{1}{2}\hbar\omega g_l(\omega)d\omega
\end{equation}

\noindent where $\omega_L=\frac{2\pi c}{L}$ is the longitudinal frequency in the system that corresponds to the shortest wavelength governed by $L$ as discussed above and $g_l(\omega)=\frac{3\omega^2}{\omega_{\rm max}^3}$ is the density of longitudinal states per mode. This gives $E_0^l$ per mode as $E_0^l=\frac{3}{8}\hbar\omega_{\rm max}\left(\frac{\omega_L}{\omega_{\rm max}}\right)^4$. $\frac{\omega_L}{\omega_{\rm max}}$ can be written as $\frac{l_{\rm P}}{L}$, where $l_{\rm P}$ corresponds to the shortest (tentatively Planck) distance. This gives:

\begin{equation}
\begin{aligned}
&E_0^l=\frac{3}{8}\hbar\omega_{\rm max}\left(\frac{\omega_L}{\omega_{\rm max}}\right)^4=\frac{3}{8}\hbar\omega_{\rm max}\left(\frac{l_{\rm P}}{L}\right)^4, \omega_L<\omega_{\rm max}\\
&E_0^l=\frac{3}{8}\hbar\omega_{\rm max},\omega_L=\omega_{\rm max}
\label{zero2}
\end{aligned}
\end{equation}

Eq. (\ref{zero2}) is readily generalized for the $n$-component field. Using the normalized density of states for the $n$-component field $g_l(\omega)=\frac{n\omega^{n-1}}{\omega_{\rm max}^n}$ \cite{annals} gives

\begin{equation}
E_0^l=\frac{n}{2(n+1)}\hbar\omega_{\rm max}\left(\frac{l_{\rm P}}{L}\right)^{n+1}
\label{zero3}
\end{equation}

The vacuum energy $E_0^l$ approaches zero in the gas-like regime of field dynamics where only longitudinal oscillatory modes are operative. If only long-wavelength oscillatory modes are present in the field, $L$ is large, and the vacuum energy can be arbitrarily small according to Eq. (\ref{zero2}) (the upper limit of $L$ is given by the system size), asymptotically approaching zero and remaining positive as shown in Figure 2a.

\begin{figure}
\begin{center}
{\scalebox{0.6}{\includegraphics{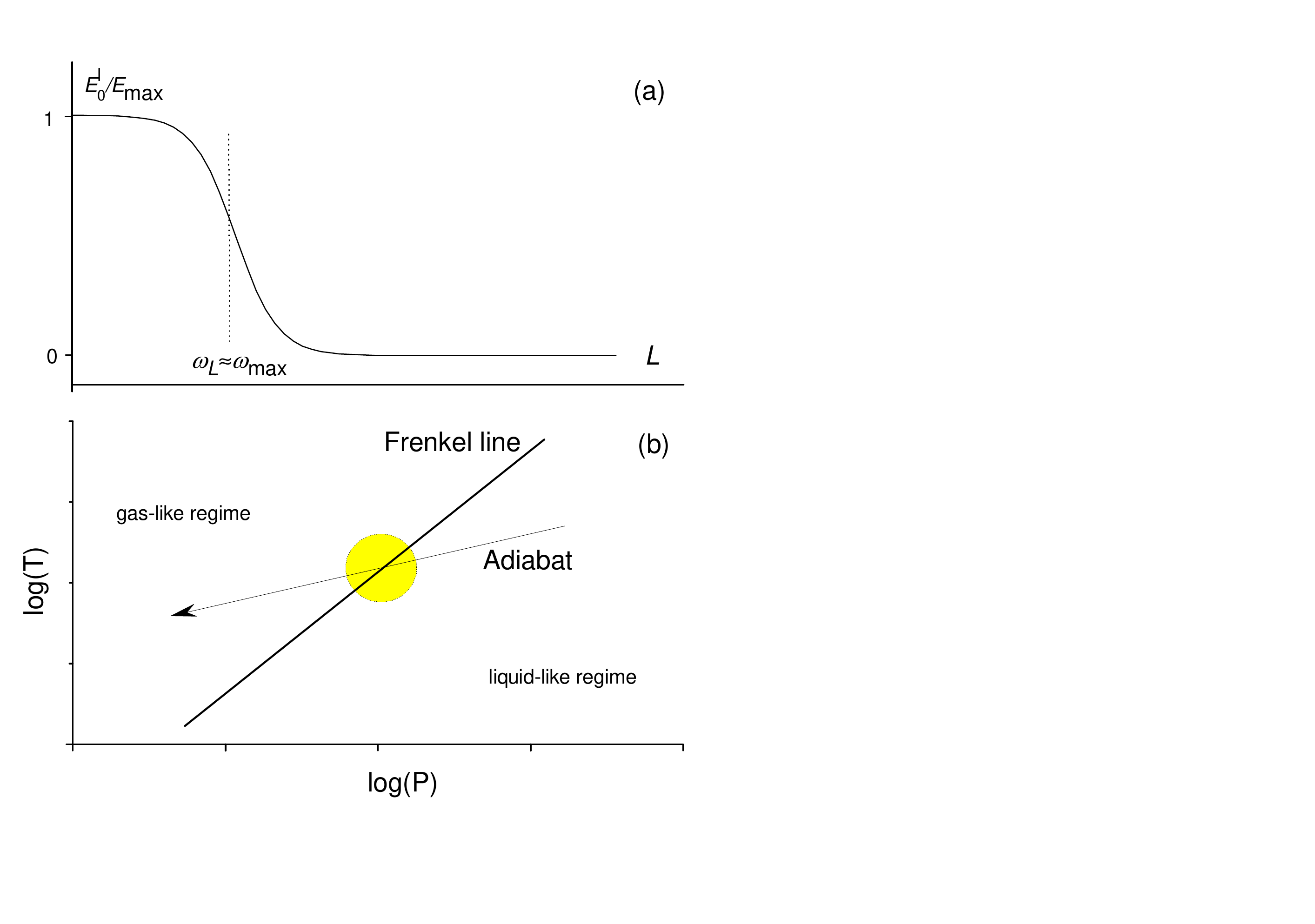}}}
\end{center}
\caption{(a) Schematic plot of (a) relative vacuum energy $\frac{E_0^l}{E_{\rm max}}$ as a function of $L$; and (b) the Frenkel line separating the liquid-like from the gas-like regime of field dynamics. The arrowed line, the adiabat of the current Universe expansion, crosses the Frenkel line, resulting in the crossover of the field dynamics from liquid-like to gas-like where $L\gg l_{\rm P}$. The shaded circle corresponds to $\omega_{\rm F}\approx\omega_{\rm max}$ and $\omega_L\approx\omega_{\rm max}$ ($L\approx l_{\rm P}$), the part of the phase diagram where the field vacuum energy decreases from its maximum to a small value that asymptotically reaches zero in the late epoch.}
\label{illustr}
\end{figure}

Eqs. (\ref{zero1},\ref{zero2},\ref{zero3}) highlight the important result of this work: the vacuum energy of the field moving in a double-well (or multi-well) potential can acquire any value between its maximum and zero.

Here, we have considered a scalar field $\phi$. Our treatment requires the presence of field interaction in the double-well form only. Therefore, the results are applicable to other appropriately interacting fields with the anharmonic interaction potential considered here.

\section{Discussion of possible cosmological implications}

Let us consider $\phi$ in Eq. (\ref{field}) as the cosmological scalar field giving rise to the inflation and assume a commonly discussed case study where the cosmological constant $\Lambda$ is related to the vacuum energy of the scalar field \cite{carroll,linde,liddle}. We consider a generic form of potential $V(\phi)$ in the double-well form in Figure \ref{wells}, the consideration sufficient for our mechanism to work. Specifying the precise form of $V(\phi)$ is not required at this stage and can include additional constraints related to observations \cite{liddle}. For the purpose of discussing the evolution of the field modes, we continue to consider the field quantized in Minkowski spacetime and do not account for the Hubble damping term. As follows from the discussion below, the important modes under discussion involve wavelengths $L$ which are sufficiently short not to be affected by the space curvature and to be within the horizon.

Our previous discussion provides a simple physical mechanism for the evolution of the vacuum energy of the scalar field. To explain the experimental data, the theory should include two parts \cite{carroll,adams,linde,liddle}: (a) inflation is large in the early epoch of the Universe and (b) due to some mechanism, inflation slows down to the currently observed small value. Our discussion provides a simple physical mechanism for both parts, and further predicts that the vacuum energy will be slowly decreasing towards zero, as follows.

In the early epoch of the Universe, the dynamics of the cosmological field $\phi$ was liquid-like as discussed below, similarly to liquid dynamics at high pressure. In this state, the vacuum energy is maximal (or close to it depending on $\omega_{\rm F}$ as discussed above), implying that inflation and related acceleration are large. Importantly, the slope of the adiabat on the pressure and temperature diagram is lower than the Frenkel line separating the liquid-like and gas-like states. The slope of the Frenkel line, $\alpha$, in the logarithmic coordinates $\log(T)=C+\alpha\log(P)$ is close to 0.8, where $C$ is a constant \cite{prl}, whereas $\alpha$ on the adiabat is close to 0.4 for a weakly-interacting system. The lower slope of the adiabat as compared to the Frenkel line was also found in the recent data \cite{pre}.

Therefore, the adiabatically expanding Universe crosses the Frenkel line from below (see Figure 2b). This has two implications. First, the field $\phi$ exists in two different dynamical regimes during the Universe inflation: liquid-like regime in the early epoch and gas-like regime in the late epoch. Second, the field dynamics in the inflating Universe undergoes a well-defined {\it crossover} from liquid-like to gas-like. In the gas-like regime, only one longitudinal mode remains, and the vacuum energy is governed by Eq. (\ref{zero2}). As the distance between the adiabat and the Frenkel line in Figure 2b increases during inflation, the field dynamics becomes increasingly gas-like, with the accompanying increase of $L$. Then, Eq. (\ref{zero2}) predicts that the vacuum energy asymptotically decreases to zero.

We note that the profound change of the total vacuum energy takes place in a fairly narrow range of $\omega_{\rm F}$ and $L$. Indeed, $E_0^t$ in Eq. (\ref{zero1}) remains close to its maximal value in the entire range of $\omega_{\rm F}$ unless $\omega_{\rm F}\approx\omega_{\rm max}$. Similarly, $E_0^l$ in Eq. (\ref{zero2}) remains close to its maximal value as long as $L\approx l_{\rm P}$, and sharply decays to zero when $L\gg l_{\rm P}$. The region of transition of large to small vacuum energy is shown as a circle in Figure 2b. We note that the sharpness of transition of $E_0^l$ in terms of $L$ does not imply the sharpness in terms of temperature and density: appreciable change of $L$ requires large temperature and density (pressure) variations, especially because temperature and pressure have competing effects on $L$ as discussed above.

We therefore find that the evolution of the field $\phi$ during the adiabatic inflation starts with both transverse and longitudinal modes and maximal vacuum energy and acceleration in the early inflation-dominated epoch and, as a result of expansion, undergoes a dynamical crossover and arrives in the state with one long-wavelength longitudinal mode and asymptotically decreasing to zero vacuum energy in the late epoch. This picture therefore addresses the difficulty of finding a model of small, yet non-zero, vacuum energy that is currently observed \cite{carroll}: we attribute small non-zero vacuum energy to the presence of only long-wavelength longitudinal oscillations of the cosmological field due to large $L$. In this picture, the currently observed vacuum energy is the small tail of the previously large inflation in the early epoch.

$L$ can be estimated from Eqs. (\ref{zero2},\ref{zero3}). The vacuum energy calculated from the field theory up to the Planck scale is larger than the observed vacuum energy by a factor of $10^{123}$ \cite{basilakos}. If $l_{\rm P}$, the Planck length, is $l_{\rm P}=10^{-35}$ m, Eq. (\ref{zero2}) gives $L\approx 0.1$ mm, not far from the range of the cosmological microwave background radiation by order of magnitude. For $n=2$, Eq. (\ref{zero3}) gives $L\approx 1000$ km.

Since the Universe is currently accelerating, following the arrowed adiabat in Figure 2b, we predict that smaller vacuum energy (cosmological constant) and acceleration will be measured in the near future. However, we comment on the possibility of a cyclic Universe below.

Eqs. (\ref{zero2},\ref{zero3}) predict that $\Lambda$ decreases with expansion. At some point of this decrease, the cosmological equations may result in contraction. Contraction corresponds to the Universe reversing its motion on the adiabat in Figure 2b. The evolution of the vacuum energy is now reversed too: as the system crosses the Frenkel line from the gas-like to the liquid-like regime of field dynamics, the vacuum energy increases to its maximum value as discussed above. At this point, inflation and acceleration become large again as in the early epoch. This corresponds to the system completing one full cycle (provided we fix the start of the cycle at the beginning of large inflation), to be followed by decreasing vacuum energy and decreasing inflation in the next cycle.

In the above discussion, the state of large vacuum energy (including the state of minimal size in the cyclic Universe) does not require singularity or unusually small scales. One possible consequence of this is that molecular structures (e.g., DNA) might survive in the state of minimal size.

As a suggestion for future work and making predictions for observations, it will be interesting to solve the cosmological equations with varying vacuum energy \cite{basilakos,sola1,sola2,sola3} where this energy evolves as generally predicted here.

\section{Summary}

In summary, we pointed out that the vacuum energy of the interacting field can take any value between its maximal value and zero. When only long-wavelength longitudinal field modes are present, the vacuum energy is close to zero. Cosmologically, field $\phi$ starts with both transverse and longitudinal field modes operating and maximal vacuum energy in the early inflation-dominated epoch and, as a result of inflation, undergoes a dynamical crossover and arrives in the state with one long-wavelength longitudinal mode and small positive vacuum energy predicted to be asymptotically decreasing to zero in the late epoch. We propose that our results warrant further study, especially so in view that the closely-related concepts from condensed matter (liquids and supercritical state) have been recently experimentally tested with some rigor \cite{prb1,col-review,scirep1,phystoday,prl,natcom}.

I am grateful to T. Clifton, D. Mulryne, S. Thomas and V. V. Brazhkin for discussions.

\end{document}